\documentclass[
  journal=pasa,
  manuscript=research-paper, 
  year=2023,
  volume=37,
]{cup-journal}

\usepackage{amsmath}
\usepackage[nopatch]{microtype}
\usepackage{booktabs}
\usepackage{microtype,siunitx,booktabs}
\usepackage{amsmath}
\usepackage{amssymb}
\usepackage{multirow}
\usepackage{threeparttable}
\usepackage[percent]{overpic}
\usepackage[figuresright]{rotating}
\usepackage{subcaption}
\usepackage{graphicx}
\usepackage{hyperref}
\usepackage{float}
\usepackage{multirow}
\hypersetup{colorlinks=true,citecolor=blue,linkcolor=blue,urlcolor=blue}

\title{Recovering Intrinsic Pulsar Profiles and Scattering Parameters with a CLEAN-Based Algorithm for High-Precision Timing} 

\author{Adarsh Bathula}
\affiliation{Centre for Astrophysics and Supercomputing, Swinburne University of Technology, PO Box 218, Hawthorn, VIC 3122, Australia }
\alsoaffiliation{Department of Physical Sciences, Indian Indian Institute of Science Education and Research, Mohali, Punjab, 140306, India.}

\email[Adarsh Bathula]{abathula@swin.edu.au}

\author{M. A. Krishnakumar}
\affiliation{Max-Planck-Institut für Radioastronomie, Auf dem Hügel 69,
53121 Bonn, Germany}
 \alsoaffiliation{Fakultät für Physik, Universität Bielefeld, Postfach 100131, 33501
Bielefeld, Germany}
\alsoaffiliation{National Centre for Radio Astrophysics, Pune University Campus, Pune 411007, India}
\author{S. Jena}
\affiliation{Department of Physical Sciences, Indian Indian Institute of Science Education and Research, Mohali, Punjab, 140306, India.}

\begin{document}

\begin{abstract}
In high precision pulsar timing, the accurate recovery of intrinsic pulsar profiles and their associated scattering parameters is of paramount importance. In this paper, we present a comprehensive study focused on the retrieval of intrinsic pulsar profiles through the utilization of a CLEAN-based algorithm as described in Bhat et al. (2003). The primary objective of this study is to elucidate the capabilities of our pipeline in the context of recovering the intrinsic profiles and associated parameters, such as dispersion measure, frequency scaling index, scattering time, pulse broadening function, and time of arrival residuals.
We use simulated profiles to rigorously test and validate the efficiency of our recovery pipeline. These simulated profiles encompass single and multi-component Gaussians, designed to emulate the diverse nature of pulsar profiles. By comparing the recovered profiles and parameters to their injected values, as derived from simulations, we provide a robust evaluation of the pipeline's performance along with its drawbacks and limitations.
 
\end{abstract}

\section{Introduction}

The study of pulsars is a rapidly evolving field. The high rotational stability of pulsars allows us to use them as accurate celestial clocks to probe a wide range of time-domain phenomena
\citep{Hobbs_2019}. This is possible through the technique of pulsar timing, where the pulsar rotation is accurately tracked by precisely measuring the times of arrival (ToAs) of its pulses
\citep{2006MNRAS.372.1549E}. The arrival
times of these pulses can be measured with remarkable accuracy, making pulsar timing a
powerful tool for studying a broad range of astrophysical phenomena.

However, the propagation of radio waves from pulsars, through the interstellar medium (ISM) is affected by different phenomena like, dispersion, Faraday rotation, scattering, etc. The delay due to dispersion scales strictly with observing frequency ($\nu$) as $\nu^{-2}$, a possibility for its frequency dependence is predicted \citep{css2016,donner2019}. Scattering, in the context of radio pulses, is primarily a consequence of the inhomogeneous distribution of free electrons within the ISM. This makes the radio waves to take multiple paths through the ISM and will create an interference pattern at the observer's plane. Due to this multi-path propagation of the waves, the pulsed signal gets broadened. This pulse broadening due to scattering evolves as a function of frequency as $\tau_{sc} \propto \nu^{-\alpha}$ \citep{rickett77} where $\tau_{sc}$ is the scattering time and $\alpha$ is its frequency scaling index. For a Kolmogorov-type turbulence, the value of $\alpha$ is expected to be 4.4 while for a random distribution of irregularities in the line of sight, it is expected to be 4 \citep{rickett1990}. Observationally it was found that this is not strictly the case and that it varies well beyond these expected values \citep{lohmer2001,bhat2004,kmjm2016,geyer2017}. 

The temporal broadening of the pulsar profile has direct implications on the precision of pulsar ToAs, since the uncertainty in the ToAs, $\sigma_{ToA}$ is directly proportional to pulse width, $W$ and inversely proportional to the signal to noise ratio, $S/N$ ( $\sigma_{ToA} \propto \frac{W}{S/N}$) \citep{2004hpa..book.....L}. Hence, making accurate measurements and removal of scattering to recover the intrinsic pulse profile shape is important for high precision pulsar timing. This could enable the inclusion of low frequency observations of millisecond pulsars in the pulsar timing array (PTA) data sets \citep{singha2024}. There are multiple methods proposed to effectively remove the pulse broadening. One of the most effective methods to do this is by using cyclic spectroscopy \citep{demorest2011}. But this method requires the raw voltage data for scattering removal. Storage of this large volume of data is costly and moreover, for the old data sets, we may not have access to the voltage data to perform cyclic spectroscopy. 

In 2003, a modified version of the CLEAN algorithm was introduced in \cite{2003ApJ...584..782B}. This algorithm utilized a one-dimensional approach to the CLEAN method on Pulsar profiles for the purpose of deconvolving the scattering effects attributed to the Interstellar Medium (ISM). The algorithm realises a twofold outcome of restoration of the intrinsic pulse shape of the pulsar, as well as the determination of the scattering time ($\tau_{sc}$) of the Pulse Broadening Function (PBF).

In this paper, we discuss the implementation of this algorithm to de-scatter the pulse profiles using the above algorithm and its implications. The \textbf{C}LEAN \textbf{B}ased \textbf{A}lgorithm for \textbf{De}convolution of \textbf{S}catter broadening (CBADeS) is a powerful tool for correcting the effects due to scattering and improving the accuracy of pulsar timing measurements. The CBADeS can recover the underlying pulse shape with high fidelity for all pulse shapes, allowing for more precise measurements of the pulsar’s ToAs. In addition, the CBADeS has broader implications for our understanding of the ISM. By studying the long-term variations in scattering times and dispersion measure (DM) of pulsar signals, we can gain insights into the properties and structure of the interstellar medium. 
Recovering intrinsic profiles using CBADeS could help improve the detection threshold of PTAs as it will help include pulsars with scattering variations as well in the sample. 

In the following section, we discuss in detail about the method and its implementation that is used for recovering intrinsic profiles in CBADeS. A simulation package is used to create different types of data sets for testing the software as explained in section 3. In section 4 we discuss about the different parameters that CBADeS estimates. In section 5 a discussion on the variations in the recovered parameters with different noise characteristics are described. The effectiveness of CBADeS on the selection of different PBFs are discussed in section 6 followed by its limitations and possible future improvements in section 7. 



\section{Revisiting the pipeline with minor changes}\label{section2}

\subsection{Methodology}\label{methodology}

Generally, the observed pulse $y(t)$ can be modeled as a convolution of the intrinsic pulse
$x(t)$ with the PBF, $g(t)$ and the net instrumental resolution function, $r(t)$:
\begin{equation}
\begin{aligned}\label{eq:second}
y(t) = x(t) \otimes g(t) \otimes r(t) ,
\end{aligned}
\end{equation}

Where $r(t)$ is modeled as a square function with a
width equivalent to the bin-width of the pulsar profile.
The PBF is a function designed
to model the effects of scattering in the ISM. For Section \ref{section2} - \ref{section4} we have used the thin screen model \citep{10.1093/mnras/157.1.55,2003ApJ...584..782B} which is given as 
\begin{equation}
\begin{aligned}\label{eq:third}
g(t) = A \exp \left(- \frac{t}{\tau_{sc}} \right) U(t) ,
\end{aligned}
\end{equation}
where, $U(t)$ is a step function.

CLEANing entails the iterative subtraction of scaled copies of $g(t) \otimes r(t)$ from $y(t)$ until the residual profile reaches the noise floor, followed by the restoration of the obtained CLEAN components (CCs).
The algorithm is divided broadly into two parts: Profile decomposition and reconstruction

For \textbf{decomposition} the algorithm identifies the maximum value of $y(t)$ and records its corresponding phase (bin number) to obtain a CC. Next, the CC is multiplied by a loop gain parameter $\gamma$ to obtain $y_c$. which is then, convolved with the dirty beam $g(t) \otimes r(t)$ and subtracted from $y(t)$, generating a
residual profile denoted by $\Delta y(t)$. This residual then becomes the new $y(t)$ for the next iteration,
and the process repeats itself. The iterative loop continues until the maximum value of $y(t)$
falls below 3 times  the root mean square of the off-pulse region ( $3\sigma_{off}$) of the profile. This decomposition process can be represented mathematically by the following equation 
\begin{equation}
\begin{aligned}\label{eq:four}
\Delta y(t) = y(t) - \{y_c(t) \otimes [ g(t) \otimes r(t)]\},
\end{aligned}
\end{equation}
where,
\begin{equation}
\begin{aligned}\label{eq:five}
 y_c(t) = \gamma \{ max [y(t)]\} \delta (t-t_0) , 
\end{aligned}
\end{equation}

\begin{figure}[hbt!]
\centering
\begin{overpic}[width=0.75\linewidth]{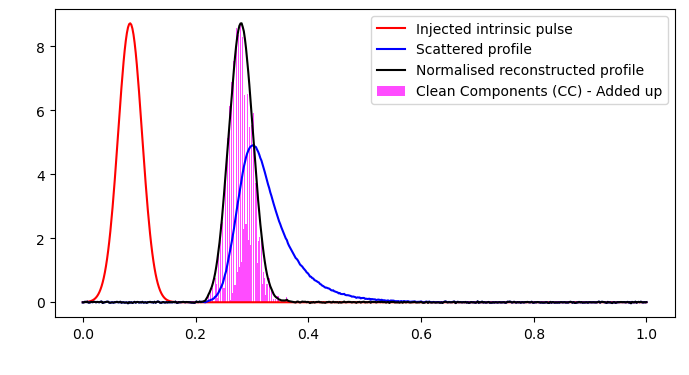}
    \put(1,15){\rotatebox{90}{\tiny Intensity (Arbitrary Units)}}
    \put(45,2){\rotatebox{0}{\tiny Pulse Phase}}
    
\end{overpic}
\caption{Plot of scattered, reconstructed and intrinsic profile along with summed up CCs. The arrangement of the CCs give the shape of the final reconstructed profile. Note that the injected intrinsic profile is shifted to the left for more visibility }
\label{CCAdd}
\end{figure}

The \textbf{reconstruction} of the intrinsic pulse profile is done by the ensemble of CCs, $C_j$ convolved with a restoring function $\rho(t)$
\begin{equation}
\begin{aligned}\label{eq:six}
y_r(t) = \sum_{j=1}^{n_c} C_j \delta(t-t_j) \otimes \rho(t), 
\end{aligned}
\end{equation}
We model the restoring function as a Gaussian distribution with amplitude and center equal to $C_j$ and $t_j$, respectively, and with width equal to the width of the profile bin.
Equation~\ref{eq:six} shows that each CC undergoes convolution with the restoring function $\rho(t)$, followed by summation to produce the profile shape. The final residual noise $\Delta y(t)$ resulting from the conclusion of the decomposition process is added to this profile. To ensure conservation of flux, the resulting profile is normalized by the area under the observed pulse $y(t)$. This process yields the final intrinsic profile of the pulsar as shown in Figure \ref{CCAdd}.

\subsection[Choosing tau\_sc using Figure of Merits]{Choosing $\tau_{sc}$ using Figure of Merits}\label{choosetau}

To find the optimal scattering timescale,$\tau_{sc}$ we conduct a search using appropriate figure of merit (FoM) criteria. To do this, we take a trial range of $\tau_{sc}$ varying from the profile bin-width to half the period of the pulsar, with a step size of the
profile bin-width. The general idea being, if the $\tau_{sc}$ is smaller than the profile bin width, the lack of profile resolution would not let us detect the scattering. On the other hand a good
upper limit is chosen to be half the period of the pulsar, since a pulsar with a scattering
timescale greater than that would essentially make the pulsar undetectable. Once the range
is set we perform the CBADeS for each $\tau_{sc}$ and then use the FoMs to determine
the best $\tau_{sc}$.
There are 5 different FoMs used in \cite{2003ApJ...584..782B},

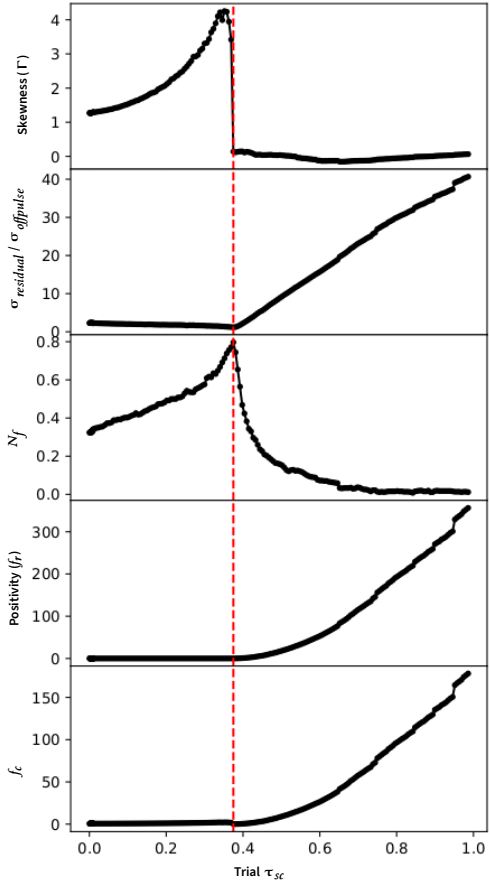
\begin{figure}[hbt!]
\centering
\begin{overpic}[width=0.75\linewidth]{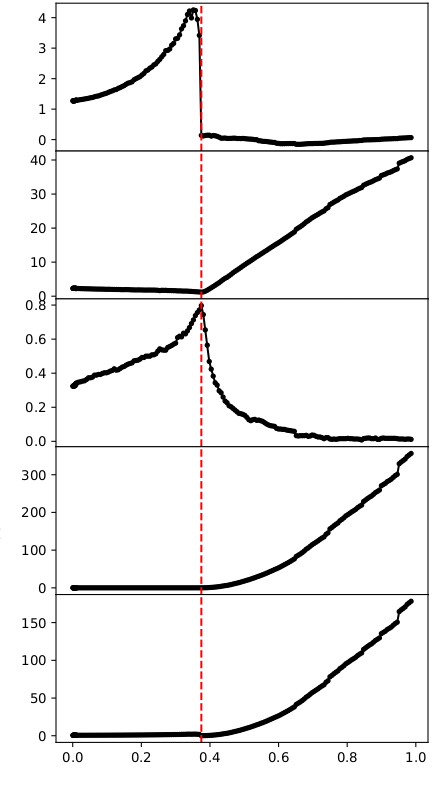}
    \put(1,85){\rotatebox{90}{\tiny \textbf{\boldmath{Skewness ($\Gamma$)}}}}
    \put(0,65){\rotatebox{90}{\tiny \textbf{\boldmath{ $\sigma_{residual}$ / $\sigma_{offpulse}$}  }}}
    \put(0,50){\rotatebox{90}{\tiny \textbf{\boldmath{$N_f$}}}}
    \put(0,30){\rotatebox{90}{\tiny \textbf{\boldmath{Positivity ($f_r$)}}}}
    \put(0,13){\rotatebox{90}{\tiny \textbf{\boldmath{ $f_c$}}}}
    \put(25,2){\rotatebox{0}{\tiny \textbf{\boldmath{ Trial $\tau_{sc}$}}}}
    
\end{overpic}
\caption{A plot of all FoMs. The red dashed line shows the best $\tau_{sc}$. }
\label{fom_1}
\end{figure}

\textbf{Minimum Residual Noise ($\sigma_{\text{residual}}$)} is defined as the ratio of the root mean square (RMS) of the final residual profile post-decomposition to the RMS of the off-pulse region of $y(t)$.

To compute the off-pulse RMS of a given profile, a discrete Fourier transform (FFT) is initially applied to the profile. Subsequently, a segment of one-sixth of the total profile is excluded from both ends, thereby isolating the central region of the profile. This process effectively excludes the signal in the profile keeping only the off-pulse region for the RMS calculation. The inverse FFT is then performed on this truncated profile reverting it to the time domain. The RMS value is subsequently derived from the standard deviation of the absolute values of the transformed data, scaled by a factor of $\sqrt{2}$. Applying the $\sqrt{2}$ scaling factor ensures that the RMS calculation reflects the true root mean square amplitude of the original signal. Without this adjustment, the RMS calculation would be affected by the doubling of amplitude during the inverse FFT operation, leading to inaccuracies in the result. 

This methodology guarantees precise determination of the off-pulse RMS, even in scenarios where the off-pulse region is exceptionally narrow or when the scattering extends on to the next pulse

\textbf{Maximum $N_{f}$} is the ratio of the number of bins in the final residual profile, post the decomposition process that satisfy
\begin{equation}
\begin{aligned}\label{eq:seven}
 | y_i - \langle y_{off} \rangle | < 2 \sigma_{rms} ,
\end{aligned}
\end{equation}
 with the total number of bins in the profile, where $y_{off}$ and $\sigma_{off}$ denote mean and RMS of the offpulse region of the profile respectively. This is different from \cite{2003ApJ...584..782B} where the criteria was to satisfy $ | y_i - \langle y_{off} \rangle | < 3 \sigma_{rms}$. This was done to get a sharper peak in $N_f$, and to clearly distinguish the $\tau_{sc}$ chosen.

\textbf{Skewness ($\Gamma$)} is defined as the asymmetry in the generated CCs post the decomposition process. If the $\tau_{sc}$ is too small or large, it results in an asymmetry in the resulting profile due to under or over subtraction. 

Following \cite{2003ApJ...584..782B},
\begin{equation}
\begin{aligned}\label{eq:eight}
\Gamma = \frac{\langle t^3 \rangle}{\langle t^2 \rangle ^{3/2}} ,
\end{aligned}
\end{equation}
where $\langle t^n \rangle$ is,
\begin{equation}
\begin{aligned}\label{eq:nine}
\langle t^n \rangle = \frac{\sum_{i=1}^{n_c} (t_i - \Bar{t})^n C_i }{\sum_{i=1}^{n_c} C_i}  ,
\end{aligned}
\end{equation}
and $\Bar{t} $ is,
\begin{equation}
\begin{aligned}\label{eq:ten}
\Bar{t} = \frac{\sum^{n_c}_{i=1} t_i C_i}{\sum_{i=1}^{n_c} C_i} ,
\end{aligned}
\end{equation}

$\tau_{sc}$ with minimal asymmetry is chosen, which manifests as a shift from left symmetric to right  as seen in Figure \ref{fom_1}

\textbf{Positivity ($f_r$)} parameter is defined as 
\begin{equation}
\begin{aligned}\label{eq:eleven}
f_r = \frac{m}{N \sigma^2_{off}} \sum_{i=1}^N [\Delta y (t_i)]^2 U_{\Delta y}, 
\end{aligned}
\end{equation}
where,
\begin{equation}
\begin{aligned}\label{eq:twelve}
U_{\Delta y} = U [-\Delta y (t) - x \sigma_{off}] , 
\end{aligned}
\end{equation}
Here, $\Delta y(t)$ is the residual noise at the end of the decomposition process. $N$ is the total number of bins. $m$
denotes a weight of order unity. The unit step function $U_{\Delta y}$ turns on when the residual $\Delta y(t)$ is significantly below
the off-pulse noise level, i.e., when an over subtraction is caused due to a $\tau_{sc}$ that is too large. The value of $x$ is set to 3/2 to define an over subtraction penalty. This makes it so that whenever a data point in the residual
noise $\Delta y(t)$ falls below $3\sigma/2$, a penalty is added. As more
number of data points fall below the threshold, the value for $f_r$ becomes greater. The point where a sharp rise begins is chosen as the best $\tau_{sc}$.

\textbf{Minimum $f_c$} is the FoM that was used to find the best $\tau_{sc}$ in \cite{2003ApJ...584..782B}. This is defined as 
\begin{equation}
\begin{aligned}\label{eq:thirteen}
f_c = \frac{f_r + \Gamma}{2} , 
\end{aligned}
\end{equation}
The trend for all five FoMs can be seen in Figure~\ref{fom_1} for a simulated profile.

\subsection{FoM selection} \label{fomselect}

\begin{table}[hbt!]
\begin{threeparttable}
\caption{FoM table}
\label{fom_table}
\begin{tabular}{lll}
\toprule
\headrow FoM & Mean  & std \\
\midrule
Residual RMS & 0.16328&0.38185\\ 
\midrule
$N_f$ & 0.19928&0.42833\\
\midrule
$f_c$ & -9.27172  & 12.68047\\
\bottomrule
\end{tabular}
\begin{tablenotes}[hang]
\item[a]units are in step size of trial $\tau_{sc}$ range. A positive mean indicates the $\tau_{sc}$ chosen is higher than the injected $\tau_{sc}$
\end{tablenotes}
\end{threeparttable}
\end{table}

\begin{figure}[hbt!]
\centering
\includegraphics[width=0.75\linewidth]{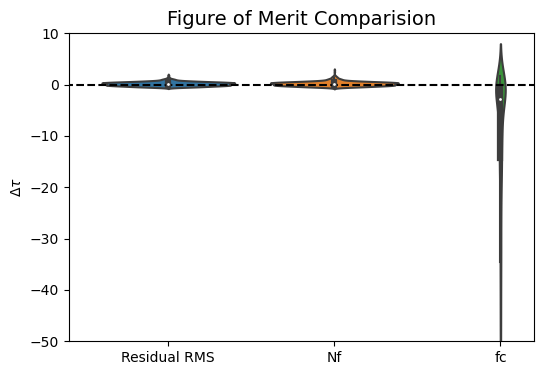}
\caption{Violin plot of the difference between injected and chosen $\tau_{sc}$ by each FoM. The units in the y axis are in terms of the step size of the Trial $\tau_{sc}$ as explained in Section \ref{choosetau}.}
\label{fomcomp1}
\end{figure}

There are four distinct FoMs ($f_c$ is derived from Skewness and $f_r$) calculated for each profile. The Skewness parameter relies on the symmetry of the summed CCs, while the other three draw insights from the characteristics of the final residual profile post the decomposition process. In certain ideal scenarios, all four FoMs converge, selecting the same $\tau_{sc}$ for a given profile. However, this convergence is not guaranteed in every case, as different FoMs might choose different $\tau_{sc}$ values, resulting in slightly varied recovered profiles.

In \cite{2003ApJ...584..782B}, the analysis involves utilizing the minima of $f_c$. However, limitations arose when dealing with the skewness parameter for various profile shapes, especially multi-component profiles.

The Skewness parameter minimises asymmetry within the CCs, which seems to falter in the context of multi-component profiles as these profiles often lack the expected symmetrical structure. The Skewness parameter tends to yield inaccurate results, thereby limiting its own and the $f_c$ parameter's effectiveness.

To assess the accuracy of these FoMs, we conducted a test using randomly generated multi-component profiles, each consisting of 512 bins. These profiles were convolved with a given PBF (in this case a single screen PBF) using a randomly chosen $\tau_{sc}$. The CBADeS was applied to each profile, and the resulting $\tau_{sc}$ values, determined by the aforementioned FoMs were compared against the injected $\tau_{sc}$. Table \ref{fom_table} and Figure \ref{fomcomp1} presents the mean difference and standard deviation of the selected $\tau_{sc}$ with the injected $\tau_{sc}$ for each FoM. A similar comparison between FoMs was conducted for single Gaussian pulses in \cite{2023arXiv230606046Y}, yielding comparable results.

As we can determine from the table \ref{fom_table} , the $\sigma_{residual}$, and the $N_f$ parameter are more accurate as compared to the $f_c$. For further analysis in this paper we have used the $\sigma_{residual}$ to determine the best $\tau_{sc}$.  
In ~\ref{appendix 1} we discuss this further with the help of a plot similar to that of Figure~\ref{fom_1}.

\section{Profile Simulators}

\subsection{Description}

Nowadays, a typical pulsar timing observation is stored in PSRFITS \citep{hotan2004} format. 
For the purpose of simulating profiles, we first choose the same single or multi-gaussian intrinsic profile for all channels without considering any frequency evolution of the profiles. The profile in each channel are then convolved with a given PBF as a function of the scattering timescale, $\tau_{sc}$. The reference $\tau_{sc}$ is given as an input at 1GHz and is then varied with the central frequency of the channel as a function of the frequency dependent spectral index $\alpha$. Once all profiles in  different channels are scattered, the profiles are dispersed using an injected DM value. Finally random noise is added with a standard deviation equal to the ratio of the peak amplitude of the profile and the desired signal to noise ratio. This data array is written into a PSRFITS format and replicates the PSRFITS files we use for Pulsar Timing data. 

\subsection{Simulation parameters}\label{sim_par}
We generated 120 PSRFITS files spanning a 10-year baseline, maintaining a consistent cadence of roughly 30 days between each epoch. Each PSRFITS file encompasses a 200 MHz bandwidth, divided into 64 channels, within the frequency range of 300 MHz to 500 MHz. The frequency dependent spectral index, $\alpha$, was kept constant at --4.0 across all epochs. Additionally, the pulsar period was selected to be approximately 3~ms divided into 512 bins, while the reference $\tau_{sc}$, was chosen to be around 0.003~ms at a frequency of 1~GHz which is $\sim$0.117~ms at 400~MHz. However, $\tau_{sc}$ is varied from epoch to epoch with some randomness. Since 400~MHz is the central frequency of our band, the results for $\tau_{sc}$ reported in this paper will be at this frequency.  A power-law model variation is injected in both $\tau_{sc}$ and DM across the 10-year time scale.

\section{Results}\label{section4}

\subsection{Profile Shape}
A fundamental aim of the CBADeS is to recover the intrinsic shape of the profile. This algorithm's distinctive advantage lies in its lack of requirements concerning the profile's intrinsic shape. Its efficacy extends to successful deconvolution of the provided PBF from  a single, multi-component, or a profile with inter-pulses. 

\begin{figure}[hbt!]
\centering
\begin{overpic}[width=0.85\linewidth]{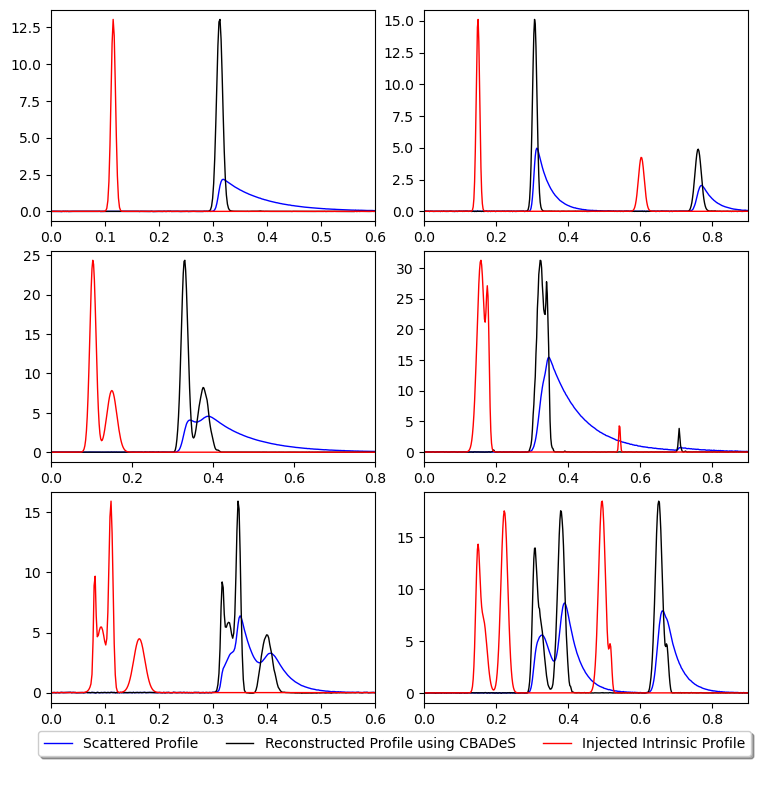}
    \put(0,45){\rotatebox{90}{\tiny Intensity (Arbitrary Units)}}
    \put(44,2){\rotatebox{0}{\tiny Pulse Phase}}
\end{overpic}
\caption{The scattered, reconstructed and injected intrinsic profiles are plotted for 6 profile shapes. Note: The injected intrinsic profile (red) is shifted to the left to prevent overlap with the reconstructed profile. }
\label{reconprof}
\end{figure}

Thorough testing of the algorithm encompassed a wide range of profile shapes, confirming its strength and versatility in handling various pulsar profile shapes. The validation process involved successfully recovering the original intrinsic profiles before convolving them with a scattering model, as demonstrated in Figure \ref{reconprof}.
It is interesting to see that CBADeS can recognise components which are not so obvious when seen in the scattered profile. Second plot in the right panel and the bottom-most plots of Figure \ref{reconprof} show this feature clearly.

\subsection[Scattering timescale tau\_sc and frequency dependent spectral index, alpha]{Scattering timescale $\tau_{sc}$ and frequency dependent spectral Index, $\alpha$}

\begin{figure}[hbt!]
\centering
\includegraphics[width=1.0\linewidth]{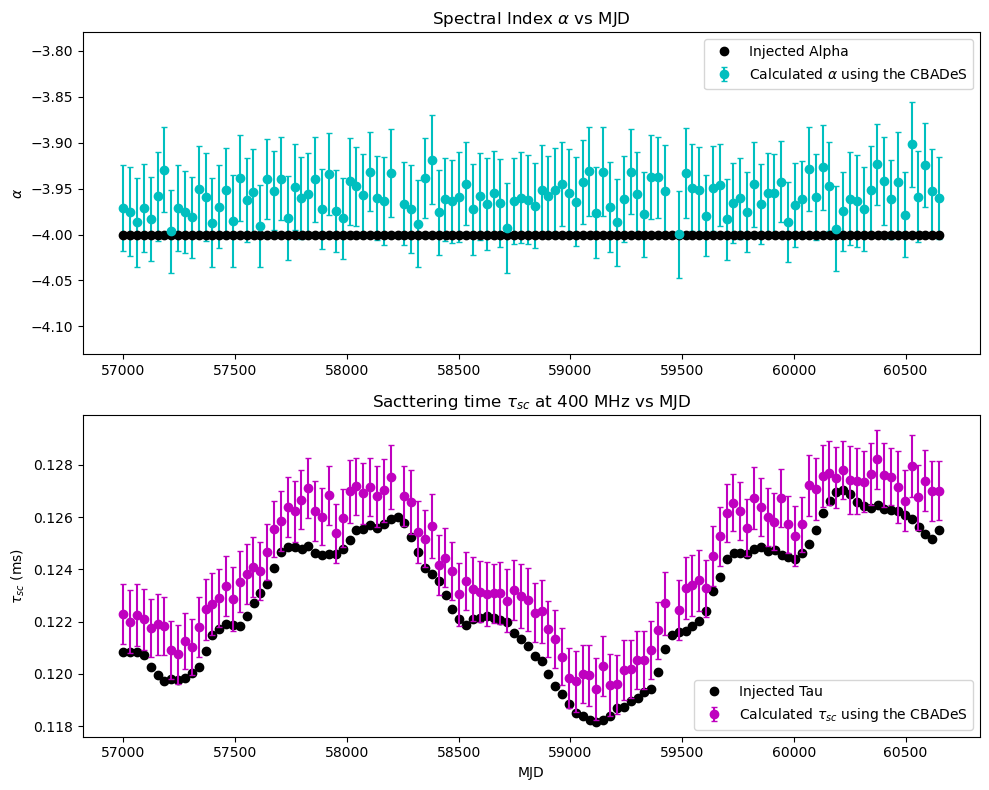}
\caption{Top Panel:Obtained Alpha values of 120 archive files. Bottom Panel: obtained $\tau_{sc}$ values at the central frequency (400MHz).}
\label{alphatau}
\end{figure}

The CBADeS is able to recover the  $\tau_{sc}$ and $\alpha$ within the error bars for high S/N profiles as shown in Figure \ref{alphatau}. 
Since, we run the CBADeS for a range of $\tau_{sc}$ the uncertainty of the measurement is chosen to be the step size of the $\tau_{sc}$ as done in \cite{2017PASP..129b4301T}.

The frequency spectral index, $\alpha$ represents the rate at which the scattering
timescale changes with frequency. It is typically described by a power-law equation,
\begin{equation}
\begin{aligned}\label{eq:fourteen}
\tau_{sc} \propto f^{\alpha} , 
\end{aligned}
\end{equation}
Since, a single simulated file comprises of 64 channels, the CBADeS was applied on each channel to obtain the optimal $\tau_{sc}$ and central frequency. Subsequently, the $\tau_{sc}$ and central frequency values underwent a Markov Chain Monte Carlo (MCMC) fitting process, employing equation \ref{eq:fourteen}, with a hundred thousand iterations to derive the parameter $\alpha$.

\subsection{Dispersion Measure}

\begin{figure}[hbt!]
\centering
\includegraphics[width=1.0\linewidth]{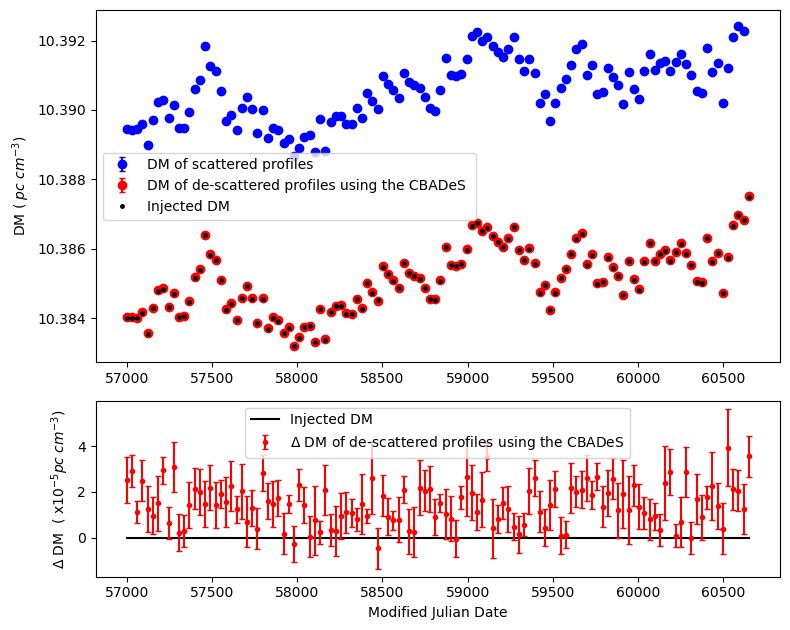}
\caption{Top Panel:The DM time-series for the scattered, de-scattered and injected DMs. Bottom Panel: Difference plot between de-scattered and injected DMs.}
\label{dmseries}
\end{figure}

To compare the dispersion measures, we run DMCalc \citep{2021A&A...651A...5K} on both, the scattered profiles as well as the recovered intrinsic profile. High S/N profiles were generated for both cases to use as a template.
Figure~\ref{dmseries} shows a plot of the DM time-series of the scattered, recovered and the injected DMs.
The \textbf{pdmp} command given in \textsc{psrchive} \citep{2004PASA...21..302H} was used to calculate the DM of the templates. These templates were then used to calculate DMs of the archive files.

There is a clear offset between the DMs of the scattered profiles and the injected DMs. The DMs obtained for the recovered profiles vary slightly but are mostly within the error bars. A slight offset could be due to the difference between the fiducial DM of the template and the original DM. 

\subsection{Time of Arrival Residuals}
\begin{figure}[hbt!]
\centering
\includegraphics[width=0.95\linewidth]{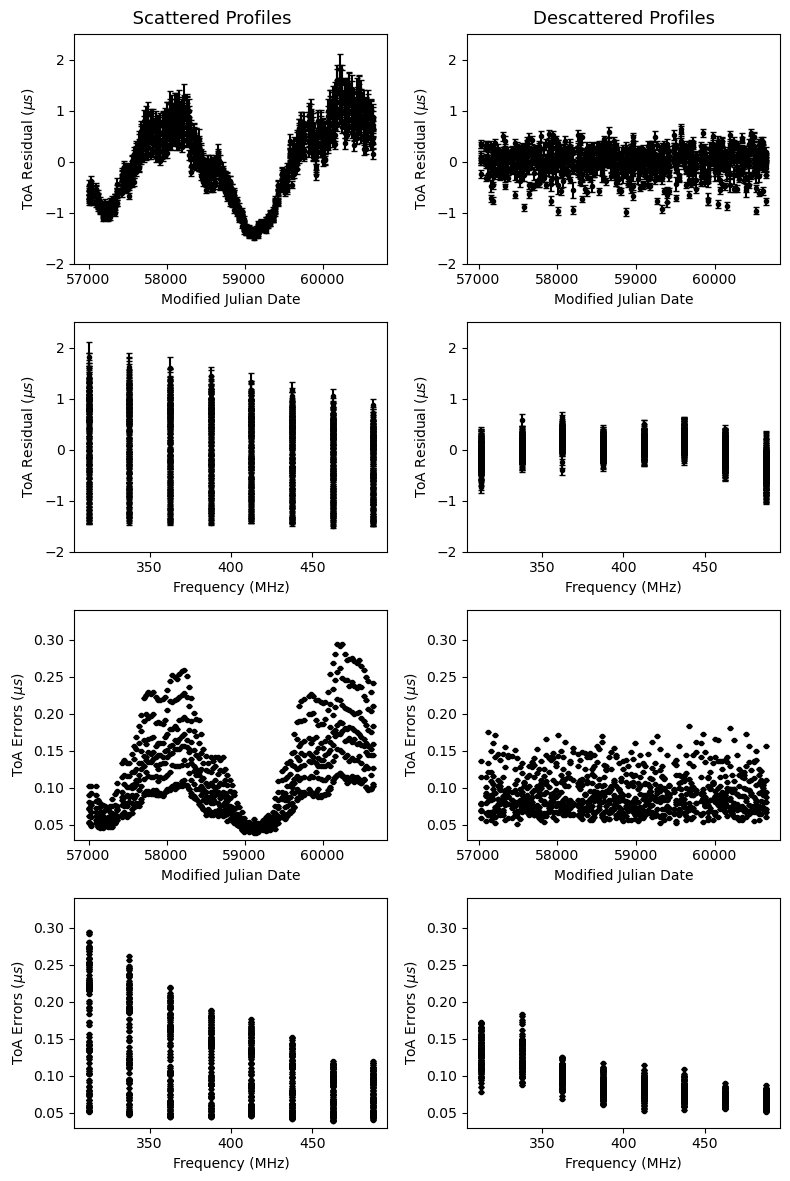}
\caption{Figure shows the obtained ToA residuals and errors for scattered profiles (Left Panel) and de-scattered profiles (Right panel) using CBADeS. ToAs and ToA errors are plotted with time and frequency for better comparison.}
\label{ToAs}
\end{figure}
ToA residuals are the difference between expected time of arrival and the observed time of arrival of the pulses. Scatter-broadening of the pulses can cause the peak of the pulse to be shifted resulting in incorrect calculations of the observed arrival time. This manifests as a variation in the ToA residuals corresponding to the variation in $\tau_{sc}$ as shown in Figure \ref{alphatau}.

ToA error is directly proportional to the width of the pulse and inversely proportional to the S/N: 
\begin{flalign} \label{fifteen2}
&\sigma_{ToA} \approx \frac{W}{S/N},&
\end{flalign}
Higher values of $\tau_{sc}$ increases the width of the pulse and subsequently the error in ToAs. CBADeS de-scatters the pulse, decreasing the width and increasing the S/N, therefore, giving us a more accurate and precise value of the ToAs.

PSRFITS files were regenerated with the same scattering variations as explained in Section \ref{sim_par}, however, no variation in DM was introduced in order to show the variation in ToAs due to scattering clearly. We compute the ToAs of the scattered and de-scattered profiles by using DMCalc \citep{2021A&A...651A...5K} scrunched to 8 ToAs per epoch. The resulting \textbf{.tim} file obtained is used to plot and compare the ToAs and their errors with respect to MJD and frequency. Figure \ref{ToAs} plots residuals and errors with MJD and frequency comparing cases for scattered and de-scattered profiles.

We see that the variation in the ToAs and its errors due to scattering in the left panel of Figure \ref{ToAs} follow the injected variation in scattering shown in Figure \ref{alphatau}. As expected, errors increase with scattering. CBADeS eliminates the scattering variation in the ToA residuals and significantly reduces the error as seen in the right panel.

\section{Variation of recovered parameters with noise}\label{snrvariance}

\begin{table*}[ht]
    \centering
    \caption{Overview of S/N dependence}
    \begin{tabular}{|l|cc|cc|cc|cc|}
        \hline
        \multirow{1}{*}{\textbf{SNR}} & \multicolumn{2}{c|}{\textbf{$\alpha$}} & \multicolumn{2}{c|}{\textbf{$\Delta\tau$ ($\mu$s)}}& \multicolumn{2}{c|}{\textbf{$\Delta$DM ( $\times10^{-5}$ $pc$ $cm^{-3}$)}} & \multicolumn{2}{c|}{\textbf{ToA Error ($\mu$s)}} \\
        \cline{2-9}
         & \textbf{Mean} & \textbf{std} & \textbf{Mean} & \textbf{std}& \textbf{Mean} & \textbf{std} & \textbf{Mean} & \textbf{std} \\
        \hline
        \textbf{SNR200} & -3.97& 0.013& 0.853 & 0.288 & 1.442& 0.902& 0.0352 &0.0126 \\
        \textbf{SNR100} &-3.91 &0.022 & 3.15 &0.396 &4.705 &0.936 &0.0555 &0.0222\\
        \textbf{SNR50} &-3.78 &0.038 &7.87 &0.877 &12.68 & 1.57&0.104 &0.0396\\
        \textbf{SNR20} &-3.47 &0.072 &22.9 &1.54 &31.17 &4.59 &0.223 &0.0718\\
        \textbf{SNR10} &-3.05 &0.105 &47.4 &3.73 &52.52 & 6.59&0.403 &0.155\\
        \textbf{SNR2} &-0.62 & 0.91& 588.1&306.6 &78.21 &65.5 &2.357 &1.038\\
        \hline
    \end{tabular}
    \label{snrtable}
\end{table*}

The CBADeS, although a highly effective tool for the deconvolution of a PBF, is strongly dependent on the S/N of the observed profile.
To assess the limits of the algorithm with respect to profile S/N, additional profiles were simulated with peak channel S/N values of 2, 10, 20, 50, 100, and 200, while keeping all other parameters identical to those used previously. These profiles were subsequently deconvolved using CBADeS, and the recovered parameters were compared against the injected values.

Table \ref{snrtable} lists the mean and standard deviation of the various recovered parameters $\alpha$, $\Delta\tau_{\mathrm{sc}}$, $\Delta\mathrm{DM}$, and ToA errors as a function of S/N, where $\Delta\tau_{\mathrm{sc}}$ and $\Delta\mathrm{DM}$ denote the differences between the parameters obtained after deconvolution with CBADeS and the injected values.

\begin{figure}[hbt!]
\centering
\begin{overpic}[width=0.95\linewidth]{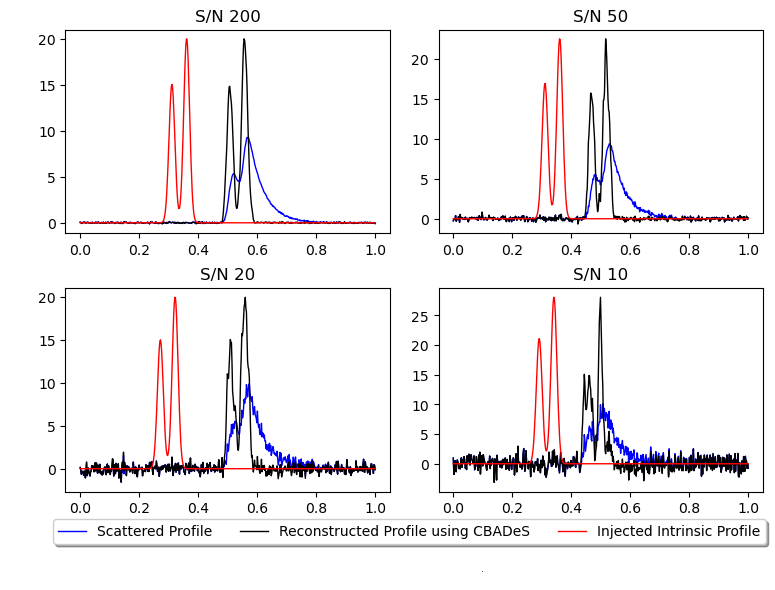}
    \put(0,30){\rotatebox{90}{\tiny Intensity (Arbitrary Units)}}
    \put(49,2){\rotatebox{0}{\tiny Pulse Phase}}
\end{overpic}
\caption{Variation in profile recovery with S/N. Note that the injected intrinsic profile is shifted to the left for visibility. }
\label{snrprof}
\end{figure}

The accuracy of intrinsic profile recovery deteriorates as the S/N decreases. Figure~\ref{snrprof} shows the same intrinsic profile convolved with the a PBF but with varying noise levels. At lower S/N, the recovered intrinsic profile deviates from the injected one, and this error can propagate into uncertainties in the obtained DM and ToA.

A clear trend is evident from Table~\ref{snrtable}, showing a systematic offset between the recovered and injected values at lower S/N, which diminishes as S/N increases. This behavior is most likely a consequence of residual scattering at lower S/N due to an elevated noise floor, which inhibits complete deconvolution. The termination criteria for the decomposition process is set to $3\sigma_{\mathrm{off}}$, as described in Section~\ref{methodology}. At low S/N, this criteria causes CBADeS to terminate early, resulting in incomplete deconvolution and consequently biasing the FoMs towards selecting a larger $\tau_{\mathrm{sc}}$.

\begin{figure}[hbt!]
\centering
\includegraphics[width=0.98\linewidth]{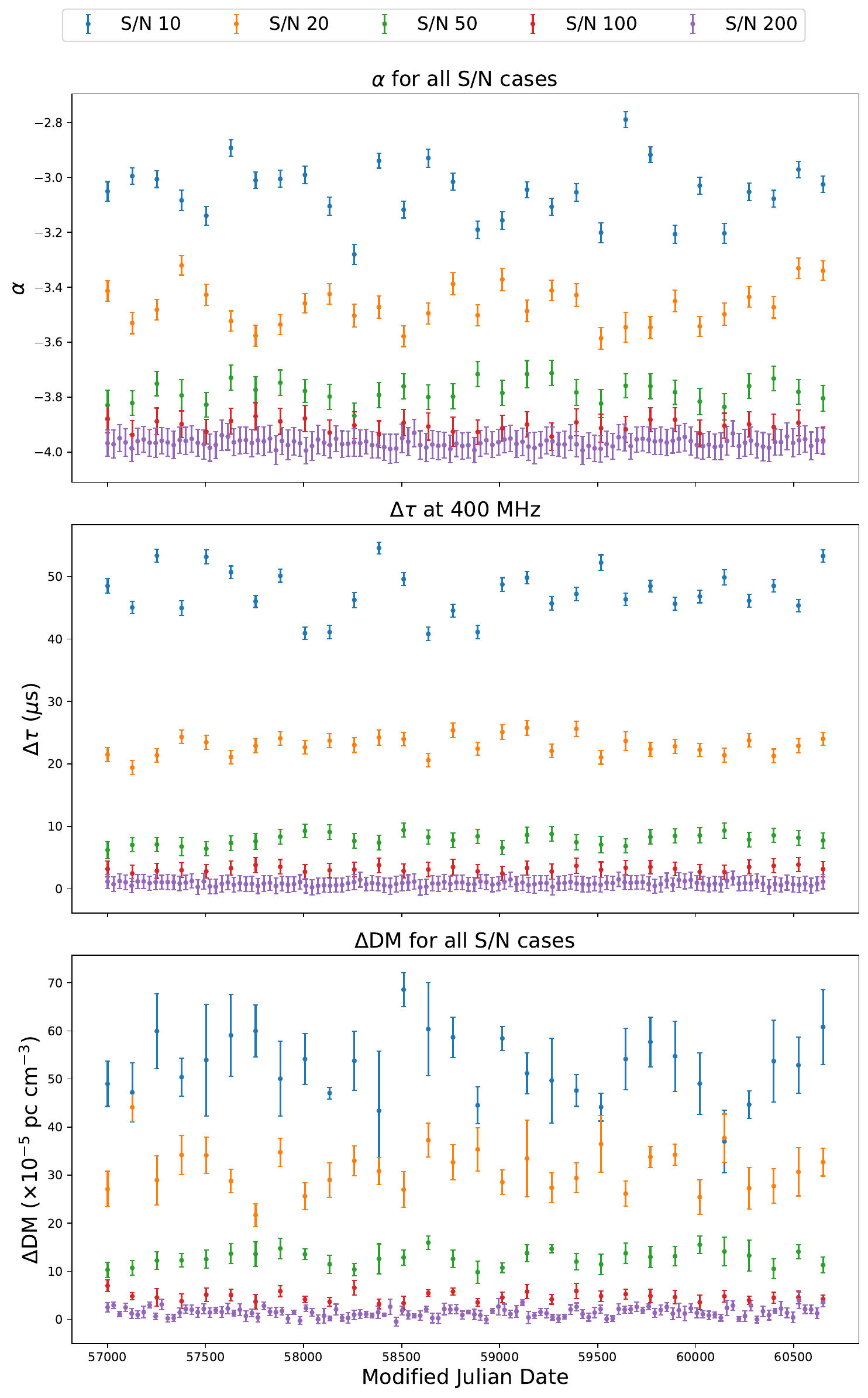}
\caption{Top panel: Variation in Alpha recovery with S/N. Injected $\alpha$ for all cases was kept constant at -4. Middle panel: Variation in $\Delta\tau_{sc}$ recovery with S/N. Bottom panel: Variation in $\Delta$DM recovery with S/N.}
\label{snralpha}
\end{figure}

The variation of $\alpha$, $\Delta\tau_{\mathrm{sc}}$, and $\Delta\mathrm{DM}$ with S/N is shown in Figure~\ref{snralpha}. It is noteworthy that the offsets in these parameters are skewed to in one direction, meaning the obtained values are always higher than the injected values as S/N decreases. As explained above, this systematic offset arises from incomplete decomposition due to the higher noise floor, affecting the recovered $\tau_{\mathrm{sc}}$ and intrinsic profile, subsequently influencing the inferred values of $\alpha$ and DM.

At a channel S/N of 2, CBADeS completely fails to recover the injected parameters, whereas at the opposite extreme (S/N = 200), the algorithm performs with excellent accuracy. 
It should be noted, however, that achieving profiles with a channel S/N as high as 200 is rare in practical observations and a  more typical S/N values lie in the range of 20 to 50. The plots for the recovery of $\tau_{sc}$, $\alpha$ and DM with variation in S/N are given below. Note that the 200 S/N case had 120 files simulated while the other S/N cases had 30 files simulated.

\section{PBF Selection}\label{pbfselection}

A major drawback in the CLEAN-based algorithm is that the recovered intrinsic profile will only be as good as the Pulse Broadening Function used to deconvolve. Currently we have been using a thin screen PBF for various simulations. Just like the thin screen model, we can also use other PBFs to deconvolve using this Algorithm. We have tested PBF2, Thick screen \citep{10.1093/mnras/157.1.55}, PBF3, Truncated Screen \citep{Cordes_2001} and PBF4, Continuous media approximation \citep{10.1093/mnras/157.1.55} as given in equations \ref{fifteen} - \ref{seventeen} for recovery of intrinsic pulse profile and timing parameters as done in Section \ref{section4}.

\begin{flalign} \label{fifteen1}
&PBF1 = A \exp \left(\frac{-t}{\tau_{sc}} \right) U(t),&
\end{flalign}
\begin{flalign} \label{fifteen}
&PBF2 = \left( \frac{\pi \tau_{sc}}{4t^3} \right)^{1/2} \exp\left(\frac{-\pi^2 \tau_{sc}}{16t}\right) U(t),&
\end{flalign}
\begin{flalign} \label{sixteen}
&PBF3 = \left(\frac{1}{\sqrt{\pi t \tau_{sc}}}\right) \exp \left( \frac{-t}{\tau_{sc}}\right) U(t),&  
\end{flalign}
\begin{flalign} \label{seventeen}
&PBF4 = \left( \frac{\pi^5 \tau_{sc}^3}{8t^5} \right)^{1/2} \exp\left(\frac{-\pi^2 \tau_{sc}}{4t}\right) U(t),&  
\end{flalign}

We further claim that the same Figure of Merits utilized in section \ref{choosetau} can serve as an effective metric for identifying the most suitable PBF.

To achieve this, the given profile undergoes deconvolution with all available PBFs, four in this case. Subsequently, the similar set of FoMs, as mentioned earlier, is employed to evaluate and compare the performance of each PBF. However, due to variations among the FoMs, it becomes imperative to scrutinize their efficacy as well.

\begin{figure}[hbt!]
\centering
\includegraphics[width=1.0\linewidth]{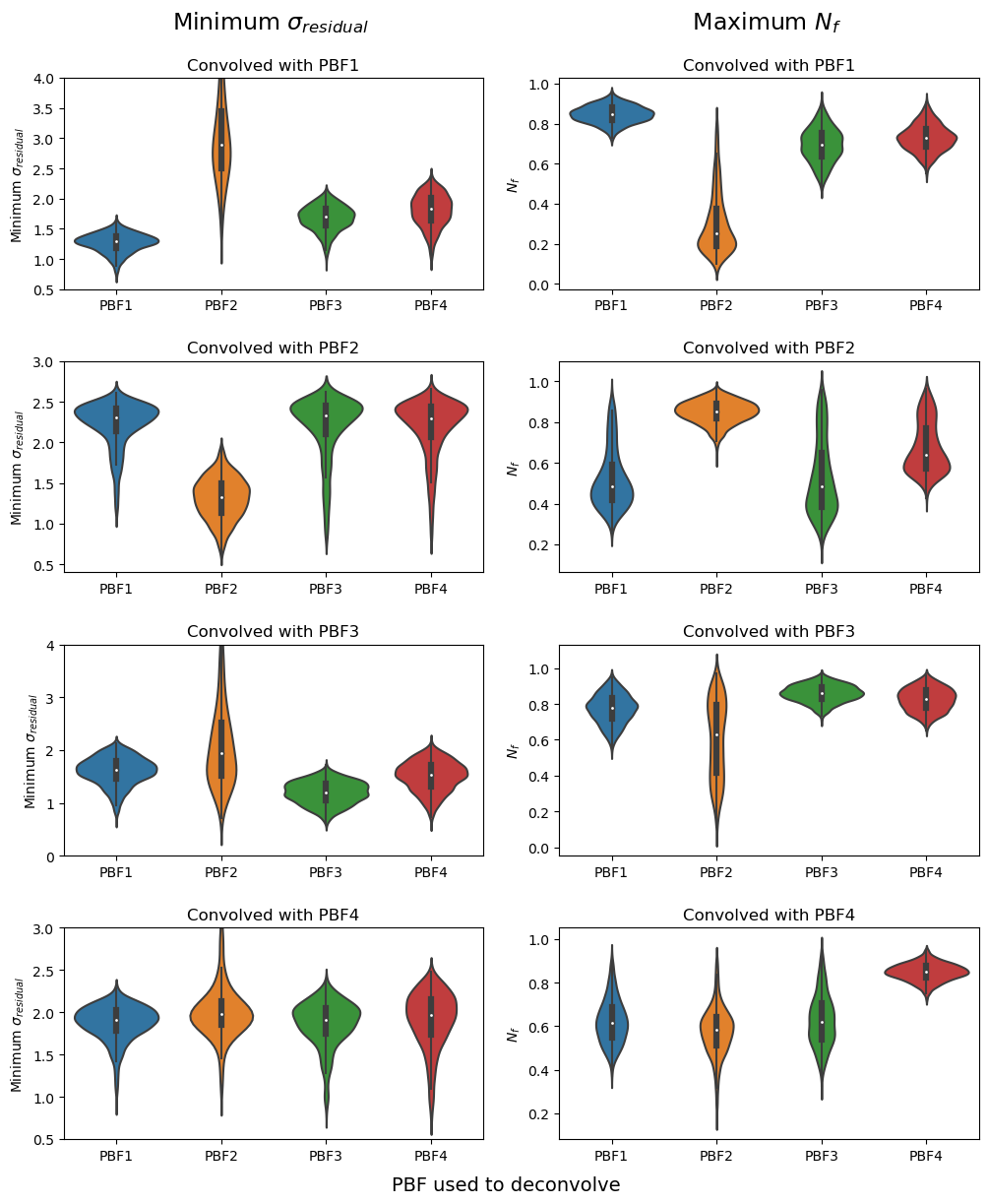}
\caption{Each panel shows FoM values (Left: minimum residual rms, Right: maximum $N_f$) when convolved with a PBF and deconvolved with all four PBFs. The PBF with the least mean is chosen for minimum residual rms and greatest mean for maximum $N_f$. }
\label{pbfselecttest}
\end{figure}

To test the effectiveness for PBF selection via the FoM criteria, a thousand, randomly generated multi-component profiles were simulated. These profiles were then convolved with PBF1 using a randomly selected $\tau_{sc}$ value. These profiles were then deconvolved using all 4 PBFs and the minimum value for the minimum residual rms, the maximum value for the $N_f$ were noted down. 
We repeated this procedure by generating a thousand more profiles each, i.e., profiles were convolved with one of the three remaining PBFs, and then deconvolved using all four PBFs. Similar procedure was followed as done for PBF1. Violin plots for minimum residual rms, and maximum $N_f$ values are shown in Figure \ref{pbfselecttest}.

To further evaluate the performance of each PBF, we employed Bayesian modeling techniques using the PyMC3 library in Python \citep{pymc2023}. The Bayesian model was constructed to estimate the posterior probabilities of each PBF, given the observed minimal residual RMS and maximum $N_f$ values. We adopted a uniform prior distribution to represent the initial probabilities of each PBF and specified a categorical likelihood based on the observed data.

Markov Chain Monte Carlo (MCMC) sampling was performed using four chains, with 1000 tuning iterations and 1000 draw iterations per chain. The posterior samples were analyzed to obtain the posterior mean probabilities and 95$\%$ credible intervals for each PBF, providing insights into their effectiveness in deconvolving the simulated pulsar profiles. The results are summarized in Table \ref{bayestable}.

\begin{table*}[ht]
    \centering
    \caption{Posteriors and 95\% credible intervals ( $\times 10^{-2}$).}
    \begin{tabular}{|c|c|c|c|c|c|}
        \hline
        \multirow{1}{*}{\begin{tabular}{c}Deconvolved\end{tabular}} & \multicolumn{4}{c|}{Convolved with} \\ \cline{2-5}
with& PBF 1 & PBF 2 & PBF 3 & PBF 4 \\
        \hline
        \multirow{1}{*}{PBF 1  \hspace{2mm} $\sigma_{residual}$}&99.694 (99.362 - 99.958)  & 0.1004 (0.0000436 - 0.303) &1.391 (0.718 - 2.127)& 14.540 (12.413 - 16.768) \\
        \multirow{1}{*}{}&  &  & & \\
       \hspace{5mm} $N_f$  &  99.003 (98.379 - 99.564) & 0.202 (0.00323 - 0.480) & 2.588 (1.636 - 3.598) & 0.0997(0.00324 - 0.299) \\
       
        \multirow{1}{*}{}&  &  & & \\
        \hline
        
        \multirow{1}{*}{PBF 2  \hspace{2mm} $\sigma_{residual}$}&  0.101 (0.000331 - 0.293) & 99.597 (99.190 - 99.916) & 0.100 (0.039 - 0.302)&40.134 (37.009 - 43.123) \\
        \multirow{1}{*}{}&  &  & & \\
       \hspace{5mm} $N_f$  & 0.101 (0.004 - 0.307) & 92.822 (91.161 - 94.378) & 3.083 (2.006 - 4.158) & 0.102 (0.00168 - 0.307) \\
        
        \multirow{1}{*}{}&  &  & &  \\
        \hline
        
        \multirow{1}{*}{PBF 3  \hspace{2mm} $\sigma_{residual}$}&0.102 (0.000669 - 0.314) &  0.101 (0.0000731 - 0.303) &94.523 (93.126 - 95.905)&  19.626 (17.120 - 22.065) \\
        \multirow{1}{*}{}&  &  & & \\
       \hspace{5mm} $N_f$  & 0.099 (0.004 - 0.301) & 1.296 (0.647 - 2.023) & 76.594 (73.923 - 79.183) & 1.195 (0.568 - 1.880) \\
        
        \multirow{1}{*}{}&  &  & & \\
        \hline
        
        \multirow{1}{*}{PBF 4  \hspace{2mm} $\sigma_{residual}$}& 0.102 (0.000480 - 0.300) & 0.202 (0.00572 - 0.475) & 3.985 (2.813 - 5.199)& 25.700 (22.966 - 28.381) \\
        \multirow{1}{*}{}&  &  & & \\
       \hspace{5mm} $N_f$  & 0.797 (0.294 - 1.354) &  5.679 (4.239 - 7.109) & 17.735 (15.364 - 20.045) & 98.604 (97.878 - 99.295) \\
        
        \multirow{1}{*}{}&  &  & &  \\
        \hline
        
    \end{tabular}
    \label{bayestable}
\end{table*}

We see from Figure \ref{pbfselecttest} that the mean of maximum $N_f$ (right panel) is always higher when we deconvolve the profile with the same PBF as the one it was convolved with. Similarly, the mean of the minimum residual rms (left panel) is lower for the PBF used to convolve the profile with.

For this paper, the criterion for selection is the minimum residual noise across all channels. By comparing these values across all channels for each PBF, the choice is made in favor of the PBF that results in the least mean residual noise value.



\section{Possible applications of CBADeS to real data}\label{applications}
%

The CBADeS algorithm is fundamentally designed to deconvolve the effects of scattering and recover the intrinsic pulse shape of temporally broadened radio transients from an intensity-time profile resolved at different frequency channels. Since scattering in the ISM affects various radio sources in a similar manner, the algorithm is broadly applicable to a wide range of time domain astrophysical objects, including pulsars, Fast Radio Bursts (FRBs), Rotating Radio Transients (RRATs), and magnetar radio bursts. Its model-independent approach to pulse deconvolution doesn't need to place any assumptions on the intrinsic pulse profile. this method can be used for all profiles including multi-componented profiles and profiles with inter-pulse and sub-structure. This, along with the ability to differentiate between various PBFs, makes it a versatile and powerful tool for studying both emission properties and propagation effects in the ISM.

For pulsars, particularly those observed along highly scattered lines of sight or at low radio frequencies, recovering the intrinsic pulse profile is crucial for accurate timing and polarimetric studies. Scattering distorts both the shape and the arrival time of the pulse, leading to systematic biases in measured DMs and ToAs. By accurately de-scattering the observed pulse profile, CBADeS enables reconstruction of the intrinsic profile, improving timing precision and mitigating scattering induced timing noise. In the case of giant pulses or highly variable emission, such as those from the Crab pulsar, where fine temporal features are often smeared out by scattering, the algorithm can be particularly valuable. Recovering the true intrinsic profile of these bursts would provide insights into the underlying emission mechanisms and hints about the emitting region.

In the context of FRBs, the ability to recover the intrinsic burst profile is of significant scientific importance as well. The intrinsic morphology of FRB pulses, whether single-peaked, multi-component, or temporally modulated, carries valuable hints about the progenitor system and emission mechanism. Scattering not only distorts these intrinsic features but can also bias the measurement of frequency drifts, spectral structures, and sub-burst durations. Applying CBADeS to FRB profiles allows for systematic de-scattering of the burst, improving the accuracy of subsequent analyses.

The algorithm can also be extended to other radio sources, such as RRATs and magnetar radio bursts. In these cases, where the pulse morphology and emission mechanisms are often less well constrained, CBADeS can help recover the intrinsic profile from propagation-induced effects. This, in turn, enables meaningful comparisons of emission mechanisms across different source classes. A particularly valuable feature of the algorithm is its ability to test and differentiate between multiple PBFs, thereby identifying the most appropriate scattering model for a given source. CBADeS can also be used to systematically differentiate between scattering models across various source classes, contributing to a broader understanding of the types of dominant scattering for each source class. Although, in this work, we have restricted our analysis to four specific PBFs, the algorithm is, in principle, capable of distinguishing among a wider range of models. By comparing the FoMs across different PBFs, CBADeS can provide valuable insights into the scattering in the ISM and its turbulent structure.

Since CBADeS requires only a frequency-resolved pulse intensity array in PSRFITS format, it can be seamlessly applied on all data files saved in this format, i.e., pulsars, FRBs, RRATs, and other transient radio sources. The primary limitation of the algorithm lies in its dependence on the S/N of the observed profile. Reliable deconvolution and accurate PBF selection require sufficient S/N in each frequency channel, as low S/N profiles may lead to incomplete de-scattering. Therefore, while CBADeS is broadly applicable to any scattered profile, its optimal performance is achieved for observations with moderate to high S/N.

Beyond its immediate applications, CBADeS holds potential for several emerging research paths. These include characterizing temporal variations in scattering to probe ISM turbulence, comparing scattering models across different host environments, and studying the frequency dependence of scattering indices to test deviations from Kolmogorov turbulence. Its flexibility positions CBADeS as a valuable tool for current and future wideband radio observations, enabling robust and model-aware de-scattering across a diverse range of astrophysical sources.

\section{Limitations and Future Endaveours}
Although CBADeS is a powerful technique, it has its limitations. It often helps to keep them in mind while doing your analysis on pulsar data. 
\subsection{Limitations}
\subsubsection{Dependance on S/N}
As outlined in \ref{snrvariance}, the accuracy of profile recovery hinges significantly upon the signal to noise ratio. Notably, profile reconstruction can be compromised in instances of low S/N profiles.
\begin{figure}[hbt!]
\centering
\begin{overpic}[width=0.85\linewidth]{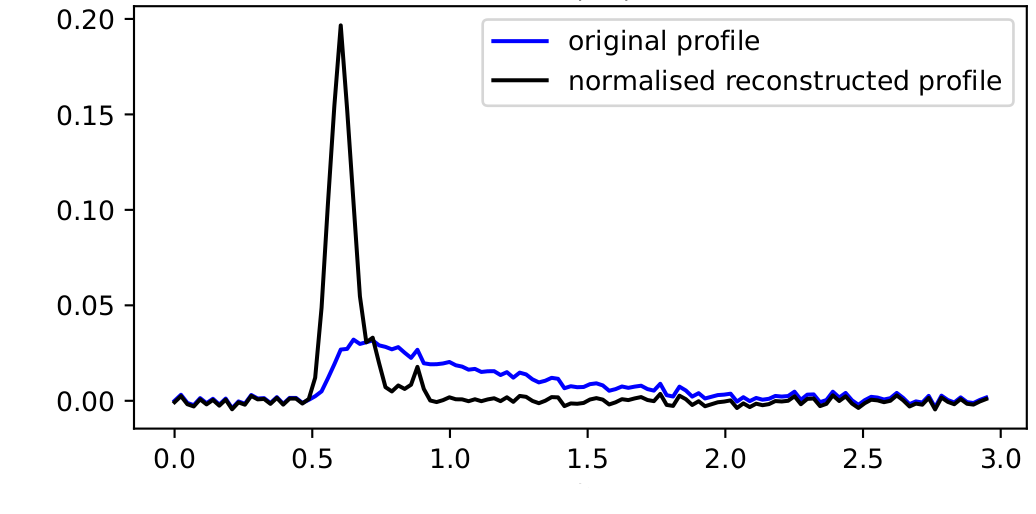}
    \put(0,15){\rotatebox{90}{\tiny Intensity (Arbitrary Units)}}
    \put(49,1){\rotatebox{0}{\tiny Time (ms)}}
\end{overpic}
\caption{An instance of a noise spike on the scattered pulse (at $\sim$ 0.9ms) being amplified on the recovered pulse. Note that the intrinsic pulse is only around 7 bins wide.}
\label{noisespike}
\end{figure}
Furthermore, owing to the techniques nature, certain instances of noise spikes within low S/N profiles might undergo amplification, potentially being misidentified as genuine components as seen in Figure \ref{noisespike}. This occurrence is particularly observable in profiles with low bin resolution. However, even post amplification, these noise components are quite small. Pulsar observations are multichannel observations, and averaging the profiles across all channels can effectively mitigate this issue, given that these noise spikes retain a random nature.
  
\subsubsection{Figure of merits}
The effectiveness of the algorithm hinges on the precision of the Figures of Merit (FoMs). The key advantage of this pipeline is its ability to operate without having to imposing constraints on either the intrinsic profile or the PBF. The pipeline conducts a thorough exploration in both parameter spaces, guided by the FoM criteria.

However, it's crucial to note that the accuracy of the deconvolution process and the resulting intrinsic pulse is contingent upon the accuracy of the PBF and in turn, are dependent on how accurately the FoMs select the $\tau_{sc}$. Any discrepancy in this determination can lead to a source of error in the overall process.

\subsubsection{distribution of CCs}
The reconstruction of wide intrinsic pulses, especially in profiles with very high bin resolutions such as 1024 or 2048 bins, presents a challenge using the standard restoration function explained in section \ref{methodology}. Typically, we convolve CCs with a Gaussian, where the amplitude and center match the CC, and the width corresponds to the bin size.

However, in the case of extremely wide intrinsic pulses, the CCs may not span the entire width of these pulses, resulting in certain gaps that manifest as additional noise spikes. To address this issue, an adjustment is made to the restoration function by multiplying the Gaussian width by a factor, which, proves effective in mitigating the gaps. Tests have been conducted to identify the optimal Gaussian width for profiles with varying bin numbers.

\subsection{Future endeavours}
In the future, we aim to improve the pipeline's capabilities by incorporating additional PBFs.  This will broaden the pipeline's applicability in mitigating scattering effects in pulsar profiles as well as  improve its ability to accurately model the ISM.
Our specific focus on improving the accuracy of profile shape recovery involves refining the deconvolution process, especially in scenarios with wide profiles or low S/N observations.

Currently, processing a fits file can be time consuming. This can vary depending on the type of profile, S/N values, number of bins, number of channels, etc. A fits file with 64 channels can take approximately 5-8 minutes. Thus, we plan to optimize the algorithm's computational efficiency further to reduce processing time. 

Finally, we will apply the algorithm to a wider array of real observed pulsar datasets, to study the efficacy of the algorithm as well as any intricate properties and characteristics of the ISM. 

Through these targeted efforts, our objective is to strengthen the algorithm's versatility, computational efficiency, and precision in profile recovery, ensuring its continued effectiveness in advancing pulsar research methodologies.

\section{Conclusion and remarks}
Our paper delves into the CLEAN-based algorithm (Bhat et al., 2003), employing it for de-scattering pulsar profiles by deconvolving a PBF. We applied the algorithm to simulated pulsar profiles, covering both single and multi-Gaussian pulses, and successfully extracting key parameters like $\tau_{sc}$, $\alpha$, DMs, and ToAs.

Expanding our investigation, we explored four alternative PBFs, developing methods to judiciously select the most suitable one for a given profile using the same FoMs used in \cite{2003ApJ...584..782B}. 

Despite these achievements, we acknowledge and address inherent limitations and try improving the robustness of the algorithm. The above investigation underscores the algorithm's potential and forms a 
foundation for future de-scattering studies of pulsar data. Continued refinement is vital for addressing challenges and ensuring the algorithm's relevance in pulsar research.

\begin{acknowledgement}


This work was supported in part by the Kishore Vaigyanik Protsahan Yojana (KVPY) Fellowship, awarded by the Department of Science and Technology, Government of India. We acknowledge the members of the InPTA collaboration for their comments on the draft. We also acknowledge Bhal Chandra Joshi, Ryan Shannon, and Yashwant Gupta for their valuable comments that helped improve the manuscript.

\end{acknowledgement}





\printendnotes

\bibliography{example}

\appendix

\section{Choosing the correct scattering time}\label{appendix0}

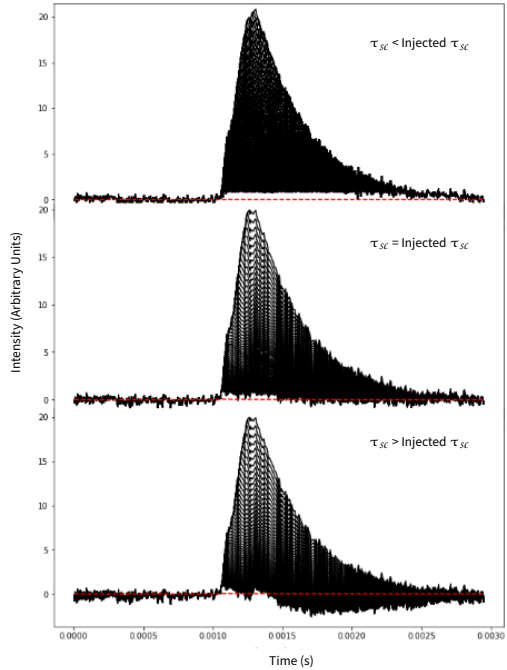
\begin{figure}[hbt!]
\centering
\begin{overpic}[width=0.75\linewidth]{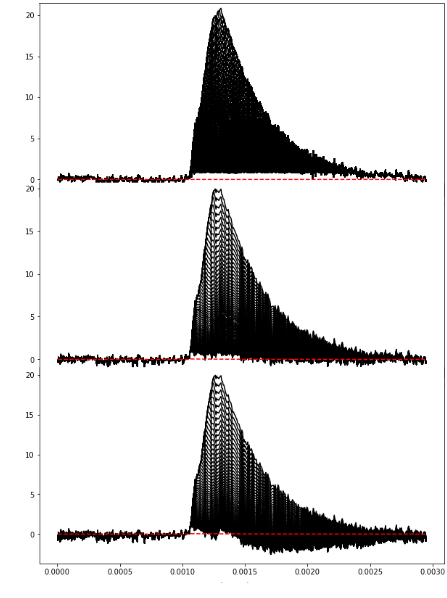}
    \put(54,93){\tiny $\tau_{sc}$ < Injected $\tau_{sc}$}
    \put(54,63){\tiny $\tau_{sc}$ = Injected $\tau_{sc}$}
    \put(54,33){\tiny $\tau_{sc}$ > Injected $\tau_{sc}$}

    \put(0,44){\rotatebox{90}{\tiny Intensity (Arbitrary Units)}}
    \put(39,0){\rotatebox{0}{\tiny Time (s)}}

\end{overpic}
\caption{The three plots show the residual profile after each iteration of subtracting the dirty beam. The three cases show how a small, accurate and a large $\tau_{sc}$ affect the final residual profile.}
\label{tauvar}
\end{figure}

It is imperative to find the correct scattering timescale for accurate subtraction of the dirty beam ($g(t) \otimes r(t)$). Choosing a $\tau_{sc}$ too small will result in under subtraction and a $\tau_{sc}$ too large would result in an over subtraction as shown in Figure \ref{tauvar}. This is the underlying concept of all the FoMs except Skewness, which relies on the arrangement of the CCs.

\section{Diverging FoMs}\label{appendix 1}

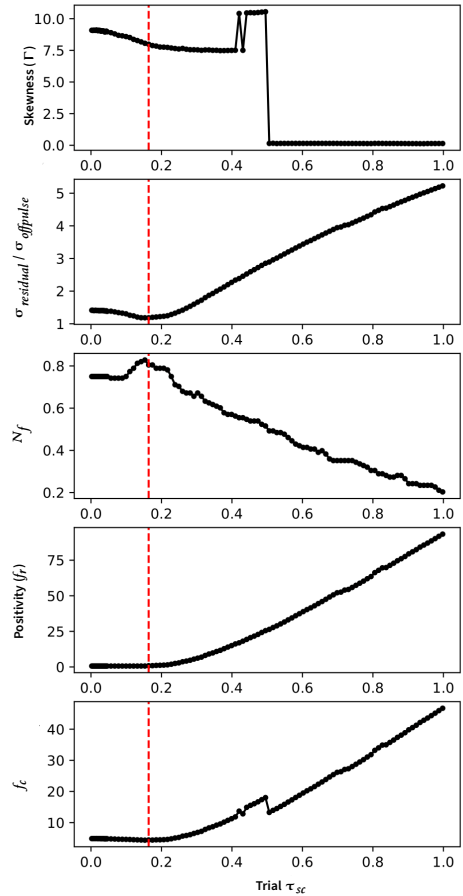
\begin{figure}[hbt!]
\centering
\begin{overpic}[width=0.75\linewidth]{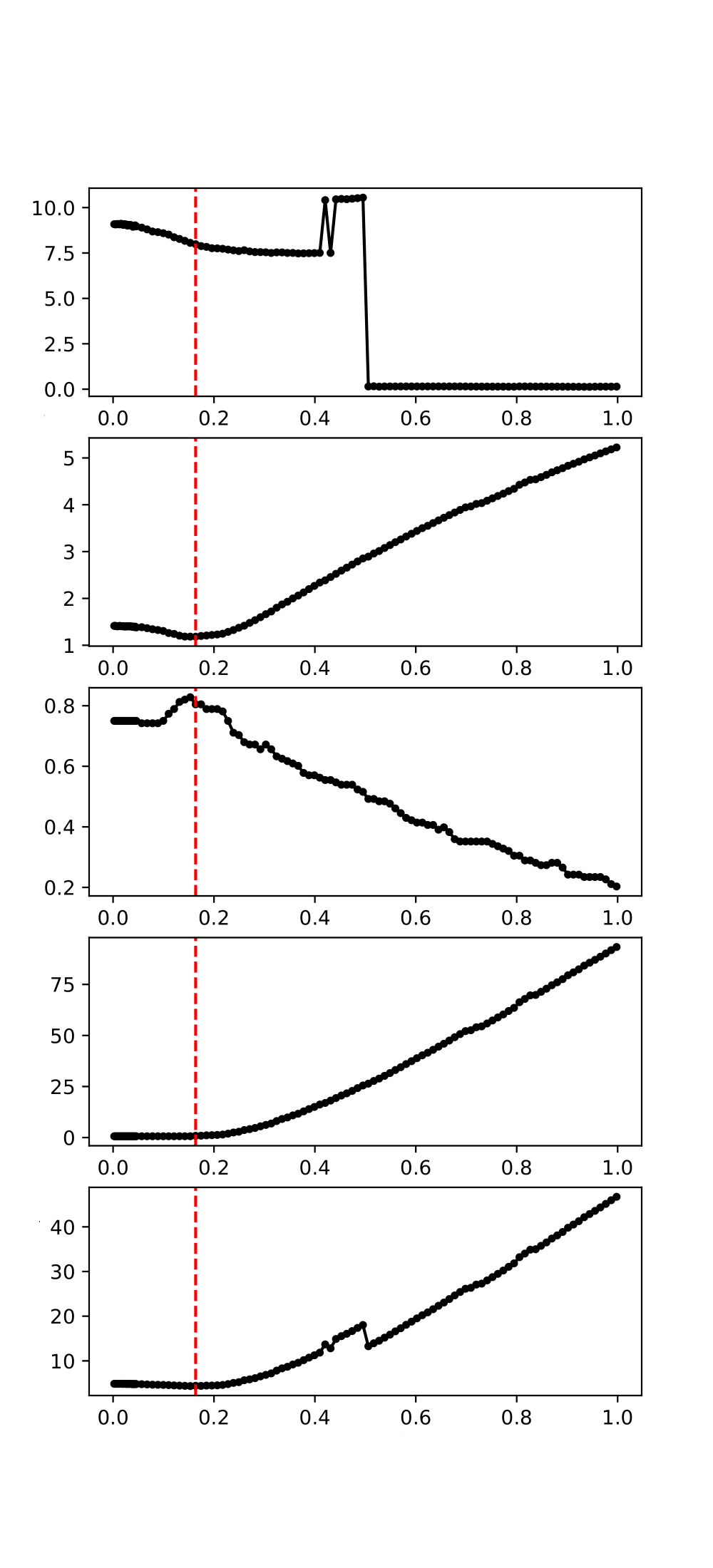}
    \put(1,77){\rotatebox{90}{\tiny \textbf{\boldmath{Skewness ($\Gamma$)}}}}
    \put(0,59){\rotatebox{90}{\tiny \textbf{\boldmath{ $\sigma_{residual}$ / $\sigma_{offpulse}$}  }}}
    \put(0,48){\rotatebox{90}{\tiny \textbf{\boldmath{$N_f$}}}}
    \put(0,30){\rotatebox{90}{\tiny \textbf{\boldmath{Positivity ($f_r$)}}}}
    \put(0,16){\rotatebox{90}{\tiny \textbf{\boldmath{ $f_c$}}}}
    \put(22,7){\rotatebox{0}{\tiny \textbf{\boldmath{ Trial $\tau_{sc}$}}}}
    
\end{overpic}
\caption{An example of FoMs showing variation in the $\tau_{sc}$ they choose. The red dashed line is the injected $\tau_{sc}$ value. We see that the Skewness parameter (Top panel) shows a drop further away from the injected $\tau_{sc}$ value resulting in a variation in the $f_c$ parameter (bottom panel) as well. }
\label{fomvariation}
\end{figure}

As detailed in Section \ref{fomselect}, the FoMs do not always converge to a single optimal \( \tau_{sc} \). This is illustrated in Figure \ref{fomvariation}, where the red dashed line represents the \( \tau_{sc} \) selected by the Minimal Residual rms method. In contrast, the Skewness parameter suggests a much higher \( \tau_{sc} \). In such instances, we prioritize the \( \tau_{sc} \) determined by the Minimal Residual rms, since it has shown to be more accurate, as evidenced in Table \ref{fom_table}.

\newpage

\end{document}